\documentclass[aps,prl,twocolumn,groupedaddress]{revtex4}
\usepackage{times}
\usepackage{graphicx}
\usepackage[amssymb,thickqspace,cdot]{SIunits}
\usepackage{amssymb,amsmath} 
\usepackage{color}
\usepackage{graphicx}
\usepackage{subfigure}
\usepackage{psfrag}

\usepackage{scrpage}
\usepackage{float}
\usepackage{longtable}
\usepackage{dsfont}
\usepackage{bbm}
\begin{document}

\preprint{}

\title{Lane reduction in driven 2d-colloidal systems through microchannels}


\author{M.\ K\"{o}ppl}
\affiliation{Universit\"at Konstanz, Fachbereich f\"ur Physik, 78457 Konstanz, Germany}
\author{P.\ Henseler}
\affiliation{Universit\"at Konstanz, Fachbereich f\"ur Physik, 78457 Konstanz, Germany}
\author{A.\ Erbe}
\email[]{Artur.Erbe@uni-konstanz.de}
\affiliation{Universit\"at Konstanz, Fachbereich f\"ur Physik, 78457 Konstanz, Germany}
\author{P.\ Nielaba}
\affiliation{Universit\"at Konstanz, Fachbereich f\"ur Physik, 78457 Konstanz, Germany}
\author{P.\ Leiderer}
\affiliation{Universit\"at Konstanz, Fachbereich f\"ur Physik, 78457 Konstanz, Germany}

\date{\today}

\begin{abstract}
  The transport behavior of a system of gravitationally driven colloidal
  particles is investigated. The particle interactions are determined by
  the superparamagnetic behavior of the particles. They can thus be arranged in
  a crystalline order by application of an external magnetic field. Therefore
  the motion of the particles through a narrow channel occurs in well-defined
  lanes. The arrangement of the particles is perturbed by diffusion and the
  motion induced by gravity. Due to these combined influences a density
  gradient forms along the direction of motion of the particles. A
  reconfiguration of the crystal is observed leading to a reduction of the
  number of lanes. In the course of the lane reduction transition a local
  melting of the quasi-crystalline phase to a disordered phase and a subsequent
  crystallization along the motion of the particles is observed. This
  transition is characterized experimentally and using Brownian dynamics (BD)
  simulations.
\end{abstract}
\pacs{}

\maketitle

Pedestrians on a walkway or in a pedestrian zone generally move in a
well-ordered fashion once the number of people per area exceeds a certain
threshold~\cite{helbing01:1068}. The reason for this behavior is that people
moving in one direction try to avoid crossing paths with people moving in the
opposite direction. This can be most easily achieved, if so-called lanes are
formed in the motion of the individuals. This phenomenon has been observed
theoretically in a large number of
systems~\cite{helbing01:1068,helbing00:1240}. Under certain circumstances it
can be shown that lane formation is also favored in many-particle
systems~\cite{dzubiella02:0214021, chakrabarti04:12401, haghooie04:061408,
haghooie05:011405, haghooie06:3601} or even in the motion of
animals~\cite{couzin02}.  Most reports of lane formation have been based on
simulations or theoretical calculations.  Only recently lane formation could be
demonstrated experimentally in a three dimensional system of oppositely charged
colloids driven by an external electric field~\cite{leunissen05:235}.

In this work we report on studies of the transport behavior of colloids in a
quasi two-dimensional (2d) setup. The colloids are superparamagnetic, therefore
the interaction energy can be continuously tuned by the application of an
external magnetic field. The particles are driven through a narrow
constriction~(channel). In addition to the analogies mentioned earlier, such a
system resembles the classical case of systems like a quantum point contact in
mesoscopic electronics~\cite{vanwees88:848,wharam88:209}. These contacts show
transport in electronic channels due to quantization effects. A classical
version of a similar scenario can be built on a liquid helium surface, which is
loaded with charges. In this system the formation of channels has been reported
as well~\cite{glasson01:176802}. The main advantage of the use of
superparamagnetic colloids instead of electrons is given by the size of the
colloids, which can be easily monitored in a standard video microscopy setup.
All relevant parameters can be gathered from the configuration data.

We compare the experimental results gained from video microscopy with
BD simulations of particle flow through constrictions under very similar
conditions. The results of both cases show qualitative and quantitative
agreement. The main focus of this discussion will be on the lane reduction
transition, giving possible scenarios for the origin of this transition and
describing the behavior of the particles in the transition region and its
close vicinity.

\begin{figure}
\includegraphics[width=\columnwidth,clip]{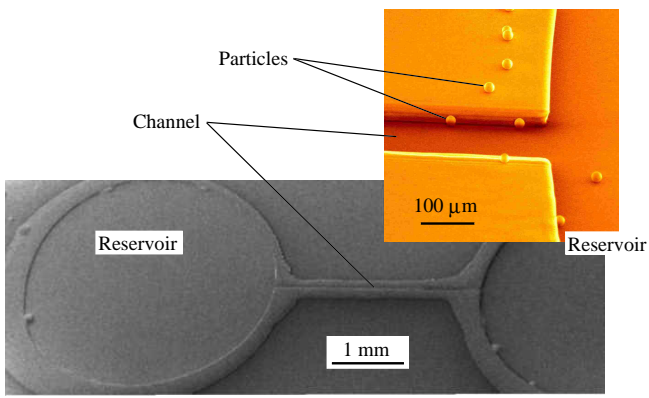}
\caption{\label{fig:setup}SEM-pictures of the channel: Shown are an overview of
the channel and a zoom to the region of the channel entrance. Some dried
particles inside and outside of the channel can be seen as well. 
}
\end{figure}

Two particle reservoirs and a connecting channel are defined on a lower
substrate using UV-lithography. SEM pictures of the channel setup and of the
channel entrance together with some dried particles inside and outside of the
channel can be seen in Fig.~\ref{fig:setup}.  The channel
(\unit{60}{\micro \meter} wide and \unit{2}{\milli \meter} long) is filled with
a suspension of superparamagnetic particles, which are commercially
available~(Dynal, particle diameters $\sigma=\unit{4.55}{\micro \meter}$,
suspended in water). Gravity confines the colloidal particles to the surface of
the channel due to the density mismatch between the particles and the liquid.
So the system is quasi two-dimensional, as long as the magnetic interactions do
not lead to out-of-plane motion of the particles. Thus for an applied uniform
magnetic field perpendicular to the monolayer the particle interaction is
purely repulsive, and its strength at distance $r_{ij}$ is given by 
\begin{equation}
V_{ij}(r_{ij}) = (\mu_0/ 4\pi) {M^2}/{r_{ij}^3}
\label{Eq:InteractionPot}
\end{equation}
with the magnetic dipole moments $M = \chi_{\mathrm{eff}} B$ of the particles,
which are proportional to  the external magnetic field~$B$.  If no boundary
conditions are imposed on the system, 2d phase transitions from a liquid phase
to the hexatic phase and finally a crystalline phase can be observed as a
function of the particle interactions~\cite{zahn99:2721}, with boundary
conditions the behavior becomes more complex~\cite{Zahn03, Bubeck99}.  In
transport through narrow channels the system orders in lanes at intermediate
strength of the interactions.  The importance of the pair-interaction can be
characterized by the dimensionless interaction strength~\cite{zahn99:2721}
\begin{equation}
\Gamma = \frac{\mu_0 M^2 \rho^{3/2}}{4\pi k_B T},
\label{Eq:PlasmaParameter}
\end{equation}
where $\rho$ denotes the number density of the particles, $k_B$ the Boltzmann
constant, and $T$ the temperature.

Even at moderate tilts of the experimental setup the particles cannot escape
the channel and gravitationally induced transport between the two reservoirs
has to take place inside the channel. At this moderate tilt the velocity of
non-interacting particles depends linearly on the tilt.  A small barrier
surrounding the channel prevents additional particles from falling into the
channel. Thus the number of particles is conserved in the system
reservoirs-channel. The video microscopy setup allows for moderate tilting
without losing optical alignment. Before starting the experiments the system is
set up completely horizontal.  We use configurations where the particles are
either all confined in one reservoir or equally distributed along the channel
and in both reservoirs. The susceptibility of the colloidal particles is
$\chi_{\mathrm{eff}} = \unit{3.07\cdot10^{-11}}{\ampere\meter^2/ \tesla}$. Thus
an external magnetic field $B=\unit{0.24}{\milli\tesla}$ corresponds to $\Gamma
\approx 2.5$.

The Brownian dynamics (BD) simulations are based on an overdamped Langevin
equation. This approach neglects hydrodynamic interactions as well as the
short-time momentum relaxation of the particles. Both approximations are fully
justified in the current experimental context. 
Typical momentum relaxation times are on the order of \unit{100}{\micro
\second} and therefore much shorter than the repetition rate of the video
microscopy setup~(\unit{10}{\second}). The colloidal trajectories
$\mathbf{r}_i(t) = (x_i(t), y_i(t))$ ($i=1,\ldots,N)$ are approximated by the
stochastic position Langevin equations with the friction constant $\xi$
\begin{equation}
\xi \frac{d\mathbf{r}_i(t)}{dt} = -
\nabla_{\mathbf{r}_i} \sum_{i\neq j} V_{ij}(r_{ij}) + \mathbf{F}_{i}^
{\mathrm{ext}} + \mathbf{\tilde{F}}_{i}(t).
\label{Eq:PositionLangevin}
\end{equation}
The right hand side includes the sum of all forces acting on each particle,
namely the particle interaction, the constant driving force $\mathbf{F}_{i}^
{\mathrm{ext}} = mg\sin(\alpha) \mathbf{\hat{x}}$ for the inclination $\alpha$
and the random forces $\mathbf{\tilde{F}}_{i}(t)$. The latter describe the
collisions of the solvent molecules with the $i$th colloidal particle and in
the simulation are given by random numbers with zero mean, 
$\langle \mathbf{\tilde{F}}_{i}(t) \rangle = 0$, and variance
$\langle \tilde{F}_{i \alpha}(t) \tilde{F}_{i \beta}(0) \rangle = 2k_B T \xi
\delta(t) \delta_{ij} \delta_{\alpha \beta}.$
The subscripts $\alpha$ and $\beta$ denote the Cartesian components.  These
position Langevin equations are integrated forward in time in a Brownian
dynamics simulation using a finite time $\Delta t$ and the technique of
Ermak~\cite{AllenTildesley,Ermak75}. Particles are confined to the channel by
ideal elastic hard walls in $y$-direction.  The channel end is realized as an
open boundary. To keep the overall number density in the channel fixed, every
time a particle leaves the end of the channel a new particle is inserted at a
random position (avoiding particle overlaps) within the first 10\% of the
channel, acting as a reservoir.

Starting from a random particle distribution within the channel, we first
calculate an equilibrium configuration ($\mathbf{F}_{i}^ {\mathrm{ext}} =0$) of
a closed channel with ideal hard walls. Afterwards we apply the external
driving force and allow the system to reorganize for $10^6$ time steps, before
we evaluate the configurations.
The time step $\Delta t = 7.5\cdot 10^{-5} t_B$ is used, with $t_B = \xi
\sigma^2/ k_B T$ being the time necessary for a particle in equilibrium to
diffuse its own diameter $\sigma$.  We choose $\xi = 3\pi \eta \sigma$, with
$\eta$ denoting the shear viscosity of the water. The simulations are done with
$2000-4500$ particles, for a channel geometry of $L_x = 800\sigma$ and $L_y =
(9-12)\sigma$, and $\Gamma$-values of $\Gamma \approx 70-950$.

\begin{figure}
\includegraphics[width=\columnwidth,clip]{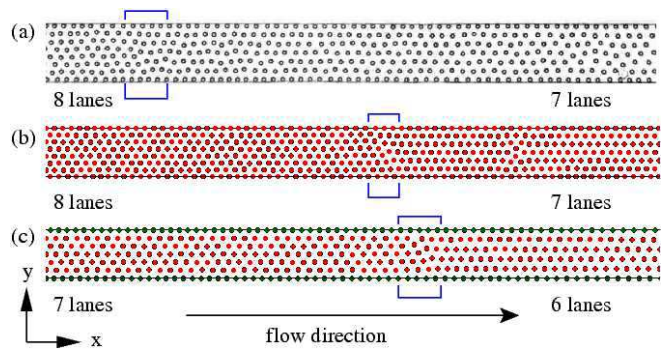}
\caption{\label{fig:channelsnapexperiment} (a) Video microscopy snapshot of
colloidal particles moving along the lithographically defined channel. (b)
Simulation snapshots for a channel ($\unit{692\,\times\,60}{\micro \meter}$, $\Gamma
\approx 2.5$) with ideal hard walls ($\unit{573.3\,\times\,45}{\micro \meter}$,
$\Gamma = 115$), (c) the same as in (b) with the particles at the walls (marked green) kept
fixed ($\unit{573.3\,\times\,45}{\micro \meter}$, $\Gamma = 902$).  The blue
rectangles mark the lane transition region.
}
\end{figure}
\begin{figure*}
\includegraphics[width=\textwidth,clip]{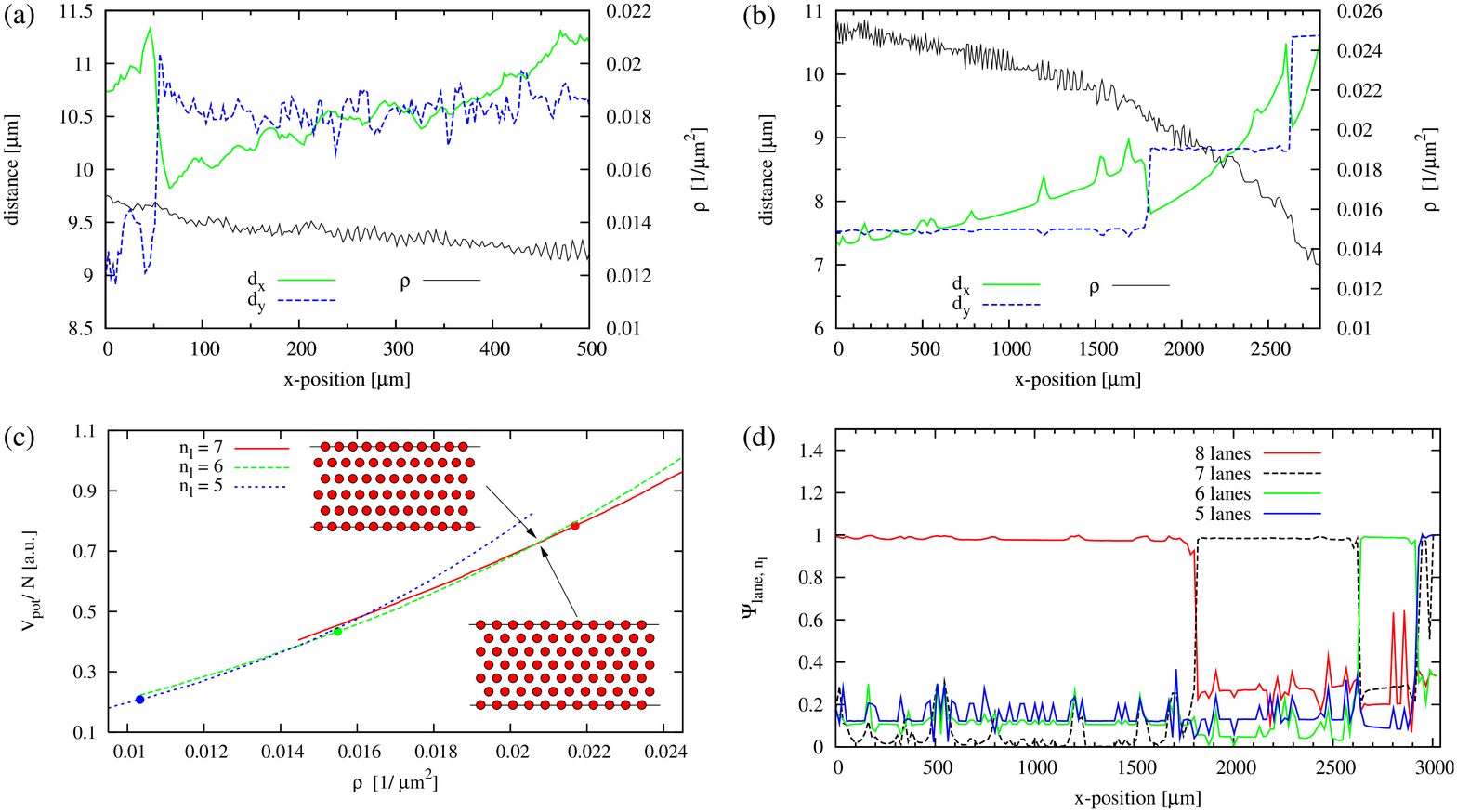}
\caption{\label{fig:densitygradient} Local lattice constants $d_x$ and $d_y$
and local particle density (a) in the experiment and (b) in the BD simulation.
The results are obtained for the systems of
Fig.~\ref{fig:channelsnapexperiment}(a) and (b) respectively. (c) Potential
energies per particle of different lane configurations as a function of the
particle density. The dots mark the perfect triangular lattices for 5, 6 and 7
lanes. Also shown are parts of the configurations with 7 and 6 lanes at the
intersection point. (d) Plots of the lane order parameter for the configuration
snapshot of Fig.~\ref{fig:channelsnapexperiment}(b). 
}
\end{figure*}
A typical snapshot from the experiment of the particles moving along the
channel is shown in Fig.~\ref{fig:channelsnapexperiment}(a). Similar snapshots
we get from simulations with co-moving
(Fig.~\ref{fig:channelsnapexperiment}(b)) and fixed boundary particles
(Fig.~\ref{fig:channelsnapexperiment}(c)), i.e., the velocity is kept to zero
for the particles at the channel wall.  In most regions of the channel the
particles are placed in a quasi-crystalline order. This behavior is due to the
strength of the particle interactions caused by the external magnetic field
(high $\Gamma$-values), which leads to quasi-crystalline behavior in unbounded
systems as well.  The formation of this order naturally gives rise to the
formation of lanes in the motion of the particles along the channel. A similar
layering phenomenon has been observed in channels under equilibrium
condition~\cite{haghooie06:3601}.  Additionally to this lane formation we
observe, both in experiment and in simulation, a decrease of the number of lanes
in the direction of motion.  In between both regions therefore a region exists in
which the particles cannot be well-ordered. This region is called the
lane-reduction zone. In Fig.~\ref{fig:channelsnapexperiment} these regions have
been marked.

The reduction of the number of lanes originates from a density gradient along
the channel. The local particle density inside the channel is shown in
Fig.~\ref{fig:densitygradient}(a) and (b) together with the particle
separations in $x$- and $y$-directions. In the experiment
(Fig.~\ref{fig:densitygradient}(a)) the density decreases monotonically along
the direction of the motion of the particles by about 20\%. The average density
in the channel shows fluctuations on the order of 10\% as a function of
time. The total increase in density, however, is less than 3\% during the total
time of the experiment.  We therefore argue that the density gradient is formed
in a quasi-static situation. This argument is confirmed by results of BD simulations
(Fig.~\ref{fig:densitygradient}(b)), where the corresponding decrease of the particle
density is observed.

The particle separations of neighboring particles in $x$- and $y$-direction are
used to calculate the local lattice constant $d$. Due to the density gradient along
the channel, the crystal is not in its equilibrium configuration at all points
along the channel. Thus the local lattice constant $d_x$, calculated from the
particle separations in $x$-direction, can deviate from the local lattice constant
$d_y$, calculated from the particle separations in $y$-direction. At the left
end of the channel, $d_x$ is larger than $d_y$, indicating that the crystal is
stretched along the $x$-axis.  At the point of the lane-reduction transition
the crystal changes back to a situation, where $d_x$ is smaller than $d_y$.
This is achieved by decreasing $d_x$ and increasing $d_y$ by about 20\%
simultaneously. Due to this behavior no non-monotonic change in the local density
can be observed at the position of the lane-reduction transition. The behavior
of the system shows that the stretching of the crystal before the transition
causes an instability towards decreasing the number of lanes. This decrease
compresses the crystal along the $x$-direction, but apparently lowers the total
energy of the system.

\begin{figure}
\includegraphics[width=\columnwidth]{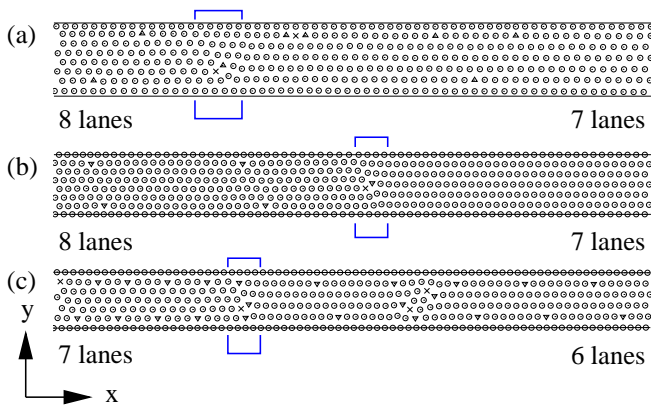}
\caption{\label{fig:voronoi} Delaunay triangulation of the particles moving in
the channel. The particles are coded according to the number of nearest
neighbors they have. Open circles mark the bulk particles with 6 nearest
neighbors and the edge particles, symbol $\times$ corresponds to fivefold
symmetry, and symbol $\triangledown$ to sevenfold symmetry.  (a) Experiment: In
order to minimize the effects of fluctuations on a short time scale, 50 images
have been averaged. (b) BD simulation for a channel with parallel walls. (c)
Equilibrium BD simulation for a channel with non-parallel walls ($\Delta L_y =
1 \sigma$ between both channel ends, $\Gamma = 28.0$).
}
\end{figure}

This can be qualitatively confirmed by the following rough estimation: Starting
from an ideal triangular configuration with a given number of lanes ($n_l$) in
a channel of fixed width, we calculated the potential energy per particle for
different particle densities by just scaling the channel length. Plots of these
energies per particle for different values of $n_l$ as function of the particle
density are shown in Fig.~\ref{fig:densitygradient}(c). They show clear
intersection points, indicating that for a stretched configuration with $n_l$
lanes in $x$-direction it can become energetically more favorable to switch to a
compressed configuration with $(n_l-1)$ lanes. Also equilibrium BD simulations
($\mathbf{F}_{i}^ {\mathrm{ext}} =0$) of closed channels with non-parallel
walls result in a density gradient with decreasing channel width and show lane
transitions. A snapshot is shown in Fig.~\ref{fig:voronoi}(c).

The region of lane-reduction can be well localized by an appropriate local order
parameter. We therefore divide the channel of width $L_y$ into several bins in
$x$-direction each containing $n_{\mathrm{bin}}$ particles and evaluate for
different number of lanes ($n_l$) the lane order parameter
\begin{equation}
\Psi_{\mathrm{lane}, n_l} = \left| \frac{1}{n_{\mathrm{bin}}}
  \sum_{j=1}^{n_{\mathrm{bin}}} e^{i \,\frac{2\pi (n_l-1)}{L_y} \,y_j}\right|,
\end{equation}
which is unity for $n_l$ particles distributed equidistantly across the channel
width starting at $y = 0$. As can be seen in Fig.~\ref{fig:densitygradient}(d),
the lane order parameter exhibits a clear discontinuity at the position of the
lane-reduction.  The width of the transition region is very sharp and usually
occurs within $2-3$ lattice constants. The local orientational order
parameter~$\Psi_6$, which is often used for 2d systems~\cite{zahn99:2721}, is
not so significant for this system, as it is very sensitive to any perturbation
of the sixfold symmetry. In addition to the energy instability the transition
requires a seed to occur, as will be demonstrated in the following discussion.

Important information on the nature of the lane-reduction transition can be
governed by the so-called Delaunay triangulation and the results are shown in
Fig.~\ref{fig:voronoi}. It reveals that the system is perfectly crystalline
left and right of the transition region. The transition is marked by a single
defect only. In addition to this, periodic defects can be seen close to the
walls. As can be observed in Fig.~\ref{fig:channelsnapexperiment} already, the
particle density in the lanes at the walls~(edge particles) is higher than for
``bulk'' particles.  Due to the long range nature of the repulsive particle
interactions it is energetically more favorable to place additional particles
in the lane at the wall than in a bulk lane~\cite{haghooie05:011405,
haghooie06:3601}. This density mismatch between edge and bulk particles causes
periodic defects, which are confined to the walls. The lane reduction is
triggered by such a defect in addition to the slowly varying particle density
along the channel, which makes the crystalline system unstable against
perturbations. The defect causes a local phase transition from crystalline to
disordered behavior.

Since the position of the lane reduction zone is mainly determined by the
density gradient, its location remains stable with time on average. A more
detailed analysis reveals, however, that the transition zone oscillates back
and forth around this average position. At the transition the driven particles
in the bulk lanes have to change the lane, causing the transition to move a
little bit in direction of the flow. A particle changing into the edge lane can
neutralize the defect of the transition locally. This causes a reconfiguration of
the crystal, which in turn gives rise to repositioning of the lane reduction
zone back to a region of higher density, i.e., back to the closest density
mismatch between bulk and edge particles.

To summarize we have shown the formations of lanes and the occurrence of a
lane-reduction transition in a 2d system of superparamagnetic colloids both
experimentally and by Brownian dynamics simulations. The lane formation is
induced by the repulsive particle interactions and the confining potential of
the channel. Due to those two factors, a crystal forms in the channel. This
crystal is stretched along the direction of the motion of the particles as a
response to the boundary conditions at both ends of the channel and the
particle interactions. Due to this stretching a lane-reduction transition
occurs in the channel, at which the crystal locally disorders. After the
transition the crystal is compressed in the direction of motion.

We gratefully acknowledge the support of the SFB 513, the SFB TR6 and the NIC,
HLRS, and SSC.

\bibliography{lanereduction.bib}

\begin{thebibliography}{17}
\expandafter\ifx\csname natexlab\endcsname\relax\def\natexlab#1{#1}\fi
\expandafter\ifx\csname bibnamefont\endcsname\relax
  \def\bibnamefont#1{#1}\fi
\expandafter\ifx\csname bibfnamefont\endcsname\relax
  \def\bibfnamefont#1{#1}\fi
\expandafter\ifx\csname citenamefont\endcsname\relax
  \def\citenamefont#1{#1}\fi
\expandafter\ifx\csname url\endcsname\relax
  \def\url#1{\texttt{#1}}\fi
\expandafter\ifx\csname urlprefix\endcsname\relax\def\urlprefix{URL }\fi
\providecommand{\bibinfo}[2]{#2}
\providecommand{\eprint}[2][]{\url{#2}}

\bibitem[{\citenamefont{Helbing}(2001)}]{helbing01:1068}
\bibinfo{author}{\bibfnamefont{D.}~\bibnamefont{Helbing}},
  \bibinfo{journal}{Rev. Mod. Phys.} \textbf{\bibinfo{volume}{73}},
  \bibinfo{pages}{1068} (\bibinfo{year}{2001}).

\bibitem[{\citenamefont{Helbing et~al.}(2000)\citenamefont{Helbing, Farkas, and
  Vicsek}}]{helbing00:1240}
\bibinfo{author}{\bibfnamefont{D.}~\bibnamefont{Helbing}},
  \bibinfo{author}{\bibfnamefont{I.}~\bibnamefont{Farkas}}, \bibnamefont{and}
  \bibinfo{author}{\bibfnamefont{T.}~\bibnamefont{Vicsek}},
  \bibinfo{journal}{Phys. Rev. Lett.} \textbf{\bibinfo{volume}{84}},
  \bibinfo{pages}{1240} (\bibinfo{year}{2000}).

\bibitem[{\citenamefont{Dzubiella et~al.}(2002)\citenamefont{Dzubiella,
  Hoffmann, and L\"owen}}]{dzubiella02:0214021}
\bibinfo{author}{\bibfnamefont{J.}~\bibnamefont{Dzubiella}},
  \bibinfo{author}{\bibfnamefont{G.}~\bibnamefont{Hoffmann}}, \bibnamefont{and}
  \bibinfo{author}{\bibfnamefont{H.}~\bibnamefont{L\"owen}},
  \bibinfo{journal}{Phys. Rev. E} \textbf{\bibinfo{volume}{65}},
  \bibinfo{pages}{0214021} (\bibinfo{year}{2002}).

\bibitem[{\citenamefont{Chakrabarti et~al.}(2004)\citenamefont{Chakrabarti,
  Dzubiella, and L\"owen}}]{chakrabarti04:12401}
\bibinfo{author}{\bibfnamefont{J.}~\bibnamefont{Chakrabarti}},
  \bibinfo{author}{\bibfnamefont{J.}~\bibnamefont{Dzubiella}},
  \bibnamefont{and} \bibinfo{author}{\bibfnamefont{H.}~\bibnamefont{L\"owen}},
  \bibinfo{journal}{Phys. Rev. E} \textbf{\bibinfo{volume}{70}},
  \bibinfo{pages}{12401} (\bibinfo{year}{2004}).

\bibitem[{\citenamefont{Haghgooie and Doyle}(2004)}]{haghooie04:061408}
\bibinfo{author}{\bibfnamefont{R.}~\bibnamefont{Haghgooie}} \bibnamefont{and}
  \bibinfo{author}{\bibfnamefont{P.}~\bibnamefont{Doyle}},
  \bibinfo{journal}{Phys. Rev. E} \textbf{\bibinfo{volume}{70}},
  \bibinfo{pages}{061408} (\bibinfo{year}{2004}).

\bibitem[{\citenamefont{Haghgooie and Doyle}(2005)}]{haghooie05:011405}
\bibinfo{author}{\bibfnamefont{R.}~\bibnamefont{Haghgooie}} \bibnamefont{and}
  \bibinfo{author}{\bibfnamefont{P.}~\bibnamefont{Doyle}},
  \bibinfo{journal}{Phys. Rev. E} \textbf{\bibinfo{volume}{72}},
  \bibinfo{pages}{011405} (\bibinfo{year}{2005}).

\bibitem[{\citenamefont{Haghgooie et~al.}(2006)\citenamefont{Haghgooie, Li, and
  Doyle}}]{haghooie06:3601}
\bibinfo{author}{\bibfnamefont{R.}~\bibnamefont{Haghgooie}},
  \bibinfo{author}{\bibfnamefont{C.}~\bibnamefont{Li}}, \bibnamefont{and}
  \bibinfo{author}{\bibfnamefont{P.}~\bibnamefont{Doyle}},
  \bibinfo{journal}{Langmuir} \textbf{\bibinfo{volume}{22}},
  \bibinfo{pages}{3601} (\bibinfo{year}{2006}).

\bibitem[{\citenamefont{Couzin and Franks}(2002)}]{couzin02}
\bibinfo{author}{\bibfnamefont{I.}~\bibnamefont{Couzin}} \bibnamefont{and}
  \bibinfo{author}{\bibfnamefont{N.}~\bibnamefont{Franks}},
  \bibinfo{journal}{Proc. R. Soc. Lond. B} \textbf{\bibinfo{volume}{270}},
  \bibinfo{pages}{139} (\bibinfo{year}{2002}).

\bibitem[{\citenamefont{Leunissen et~al.}(2005)\citenamefont{Leunissen,
  Christova, Hynninen, Royall, Campbell, Imhof, Dijkstra, v.\ Roij, and v.\
  Blaaderen}}]{leunissen05:235}
\bibinfo{author}{\bibfnamefont{M.}~\bibnamefont{Leunissen}},
  \bibinfo{author}{\bibfnamefont{C.}~\bibnamefont{Christova}},
  \bibinfo{author}{\bibfnamefont{A.-P.} \bibnamefont{Hynninen}},
  \bibinfo{author}{\bibfnamefont{C.}~\bibnamefont{Royall}},
  \bibinfo{author}{\bibfnamefont{A.}~\bibnamefont{Campbell}},
  \bibinfo{author}{\bibfnamefont{A.}~\bibnamefont{Imhof}},
  \bibinfo{author}{\bibfnamefont{M.}~\bibnamefont{Dijkstra}},
  \bibinfo{author}{\bibfnamefont{R.}~\bibnamefont{v.\ Roij}}, \bibnamefont{and}
  \bibinfo{author}{\bibfnamefont{A.}~\bibnamefont{v.\ Blaaderen}},
  \bibinfo{journal}{Nature} \textbf{\bibinfo{volume}{437}},
  \bibinfo{pages}{235} (\bibinfo{year}{2005}).

\bibitem[{\citenamefont{v.\ Wees et~al.}(1988)\citenamefont{v.\ Wees, v.\
  Houten, Beenakker, Williams, Kouwenhoven, v.d.\ Marel, and
  Foxon}}]{vanwees88:848}
\bibinfo{author}{\bibfnamefont{B.}~\bibnamefont{v.\ Wees}},
  \bibinfo{author}{\bibfnamefont{H.}~\bibnamefont{v.\ Houten}},
  \bibinfo{author}{\bibfnamefont{C.}~\bibnamefont{Beenakker}},
  \bibinfo{author}{\bibfnamefont{J.}~\bibnamefont{Williams}},
  \bibinfo{author}{\bibfnamefont{L.}~\bibnamefont{Kouwenhoven}},
  \bibinfo{author}{\bibfnamefont{D.}~\bibnamefont{v.d.\ Marel}},
  \bibnamefont{and} \bibinfo{author}{\bibfnamefont{C.}~\bibnamefont{Foxon}},
  \bibinfo{journal}{Phys. Rev. Lett.} \textbf{\bibinfo{volume}{60}},
  \bibinfo{pages}{848} (\bibinfo{year}{1988}).

\bibitem[{\citenamefont{Wharam et~al.}(1988)\citenamefont{Wharam, Thornton,
  Newbury, Pepper, Ahmed, Frost, Hasko, Peacock, Ritchie, and
  Jones}}]{wharam88:209}
\bibinfo{author}{\bibfnamefont{D.}~\bibnamefont{Wharam}},
  \bibinfo{author}{\bibfnamefont{T.}~\bibnamefont{Thornton}},
  \bibinfo{author}{\bibfnamefont{R.}~\bibnamefont{Newbury}},
  \bibinfo{author}{\bibfnamefont{M.}~\bibnamefont{Pepper}},
  \bibinfo{author}{\bibfnamefont{H.}~\bibnamefont{Ahmed}},
  \bibinfo{author}{\bibfnamefont{J.}~\bibnamefont{Frost}},
  \bibinfo{author}{\bibfnamefont{D.}~\bibnamefont{Hasko}},
  \bibinfo{author}{\bibfnamefont{D.}~\bibnamefont{Peacock}},
  \bibinfo{author}{\bibfnamefont{D.}~\bibnamefont{Ritchie}}, \bibnamefont{and}
  \bibinfo{author}{\bibfnamefont{G.}~\bibnamefont{Jones}}, \bibinfo{journal}{J.
  Phys. C.} \textbf{\bibinfo{volume}{21}}, \bibinfo{pages}{L209}
  (\bibinfo{year}{1988}).

\bibitem[{\citenamefont{Glasson et~al.}(2001)\citenamefont{Glasson, Dotsenko,
  Fozooni, Lea, Bailey, Papageorgiou, Andresen, and
  Kristensen}}]{glasson01:176802}
\bibinfo{author}{\bibfnamefont{P.}~\bibnamefont{Glasson}},
  \bibinfo{author}{\bibfnamefont{V.}~\bibnamefont{Dotsenko}},
  \bibinfo{author}{\bibfnamefont{P.}~\bibnamefont{Fozooni}},
  \bibinfo{author}{\bibfnamefont{M.}~\bibnamefont{Lea}},
  \bibinfo{author}{\bibfnamefont{W.}~\bibnamefont{Bailey}},
  \bibinfo{author}{\bibfnamefont{G.}~\bibnamefont{Papageorgiou}},
  \bibinfo{author}{\bibfnamefont{S.}~\bibnamefont{Andresen}}, \bibnamefont{and}
  \bibinfo{author}{\bibfnamefont{A.}~\bibnamefont{Kristensen}},
  \bibinfo{journal}{Phys. Rev. Lett.} \textbf{\bibinfo{volume}{87}},
  \bibinfo{pages}{176802} (\bibinfo{year}{2001}).

\bibitem[{\citenamefont{Zahn et~al.}(1999)\citenamefont{Zahn, Lenke, and
  Maret}}]{zahn99:2721}
\bibinfo{author}{\bibfnamefont{K.}~\bibnamefont{Zahn}},
  \bibinfo{author}{\bibfnamefont{R.}~\bibnamefont{Lenke}}, \bibnamefont{and}
  \bibinfo{author}{\bibfnamefont{G.}~\bibnamefont{Maret}},
  \bibinfo{journal}{Phys. Rev. Lett.} \textbf{\bibinfo{volume}{82}},
  \bibinfo{pages}{2721} (\bibinfo{year}{1999}).

\bibitem[{\citenamefont{Zahn et~al.}(2003)\citenamefont{Zahn, Wille, Maret,
  Sengupta, and Nielaba}}]{Zahn03}
\bibinfo{author}{\bibfnamefont{K.}~\bibnamefont{Zahn}},
  \bibinfo{author}{\bibfnamefont{A.}~\bibnamefont{Wille}},
  \bibinfo{author}{\bibfnamefont{G.}~\bibnamefont{Maret}},
  \bibinfo{author}{\bibfnamefont{S.}~\bibnamefont{Sengupta}}, \bibnamefont{and}
  \bibinfo{author}{\bibfnamefont{P.}~\bibnamefont{Nielaba}},
  \bibinfo{journal}{Phys. Rev. Lett.} \textbf{\bibinfo{volume}{90}},
  \bibinfo{pages}{155506} (\bibinfo{year}{2003}).

\bibitem[{\citenamefont{Bubeck et~al.}(1999)\citenamefont{Bubeck, Bechinger,
  Neser, and Leiderer}}]{Bubeck99}
\bibinfo{author}{\bibfnamefont{R.}~\bibnamefont{Bubeck}},
  \bibinfo{author}{\bibfnamefont{C.}~\bibnamefont{Bechinger}},
  \bibinfo{author}{\bibfnamefont{S.}~\bibnamefont{Neser}}, \bibnamefont{and}
  \bibinfo{author}{\bibfnamefont{P.}~\bibnamefont{Leiderer}},
  \bibinfo{journal}{Phys. Rev. Lett.} \textbf{\bibinfo{volume}{82}},
  \bibinfo{pages}{3364} (\bibinfo{year}{1999}).

\bibitem[{\citenamefont{Allen and Tildesley}(1987)}]{AllenTildesley}
\bibinfo{author}{\bibfnamefont{M.~P.} \bibnamefont{Allen}} \bibnamefont{and}
  \bibinfo{author}{\bibfnamefont{D.~J.} \bibnamefont{Tildesley}},
  \emph{\bibinfo{title}{Computer Simulation of Liquids}}
  (\bibinfo{publisher}{Oxford Science Publications}, \bibinfo{year}{1987}).

\bibitem[{\citenamefont{Ermak}(1975)}]{Ermak75}
\bibinfo{author}{\bibfnamefont{D.~L.} \bibnamefont{Ermak}},
  \bibinfo{journal}{J. Chem. Phys.} \textbf{\bibinfo{volume}{62}},
  \bibinfo{pages}{4189} (\bibinfo{year}{1975}).

\end{thebibliography}

\end{document}